# A simple efficient algorithm in frustration-free one-dimensional gapped systems

Yichen Huang (黄溢辰)[*]
*Institute for Quantum Information and Matter, California Institute of Technology, Pasadena, California 91125, USA*
(Dated: November 5, 2015)

We propose an efficient algorithm for the ground state of frustration-free one-dimensional gapped Hamiltonians. This algorithm is much simpler than the original one by Landau et al., and thus may be easily accessible to a general audience in the community. We present all the details in two pages.

Computing the ground state (energy) of local Hamiltonians is a fundamental problem in condensed matter physics and the emerging area of Hamiltonian complexity [4, 8]. In a recent remarkable paper, Landau et al. [7] proposed a randomized polynomial-time algorithm for the (unique) ground state of frustration-free one-dimensional (1D) gapped Hamiltonians. Huang [5] extended it to general 1D gapped systems. Chubb and Flammia [3] studied gapped spin chains with degenerate ground states.

This line of research is very technical. Here we significantly simplify the method in the hope that the results are easily accessible to a general audience. For this purpose, we will not use tricks that are not essential at a high level, even if they can improve the performance of the algorithm. The new ingredients of our approach allow us to get rid of many technical tools in [7]. We present all the details in two pages.

Consider a chain of $n$ spins (qu$d$its), where the local dimension $d$ of each spin is an absolute constant. Let $\mathcal{H}_i$ be the Hilbert space of the spin $i$; define $\mathcal{H}_{[i,j]} = \bigotimes_{k=i}^{j} \mathcal{H}_k$ as the Hilbert space of the spins with indices in $[i,j]$. Since the standard bra-ket notation can be cumbersome, in most but not all cases quantum states and their inner products are denoted by $\psi, \phi \ldots$ and $\langle\psi,\phi\rangle$, respectively. All states are normalized unless otherwise noted. Let $H = \sum_{i=1}^{n-1} H_i$ be a 1D Hamiltonian with $H_i$ acting on $\mathcal{H}_{[i,i+1]}$ (nearest-neighbor interaction). Suppose $H$ has a unique ground state $\Psi_0$ and a constant energy gap $\epsilon$. The goal is to find an efficient matrix product state (MPS) approximation to $\Psi_0$. The existence of such an MPS is a by-product of the proof of the area law for entanglement.

**Lemma 1** ([1]). *There exists an MPS $\Psi$ of bond dimension $n^{o(1)}$ such that $|\langle\Psi,\Psi_0\rangle| \geq 1 - n^{-\omega(1)}$.*

The best known algorithm is

**Theorem 1** ([5]). *There is a deterministic $n^{O(1)}$-time algorithm that outputs an MPS $\Psi$ such that $|\langle\Psi,\Psi_0\rangle| \geq 1 - 1/\operatorname{poly} n$.*

Suppose $H$ is frustration-free, i.e., $\Psi_0$ is in the ground space of each $H_i$. Assume without loss of generality that each $H_i$ is a projector, i.e., $H_i^2 = H_i$, so that the ground-state energy of $H$ is zero. Let $\tilde O(x) := O(x \operatorname{poly} \log x)$ hide a polylogarithmic factor. We give a simple proof of

**Theorem 2.** *There is a randomized $n^{\tilde O(1/\epsilon)}$-time algorithm that outputs an MPS $\Psi$ such that $|\langle\Psi,\Psi_0\rangle| \geq 1 - 1/\operatorname{poly} n$ with probability at least $1 - 1/\operatorname{poly} n$.*

We begin by recalling some known facts and/or tools.

**Lemma 2.** $|\langle\psi,\phi\rangle| \geq \Omega(n^{-2}/\sqrt{d})$ *with probability* $1 - O(n^{-2})$ *for two random states* $\psi, \phi \in \mathbf{C}^d$.

Fix a cut $i|i+1$ separating the spins $i$ and $i+1$.

**Lemma 3** ([2]). *A matrix product operator $A_i$ of bond dimension $2^{\tilde O(1/\epsilon)}$ can be efficiently constructed such that (i) $A_i \Psi_0 = \Psi_0$; (ii) $A_i \Psi_1 \perp \Psi_0$ and $\|A_i \Psi_1\|^2 \leq \Delta$ for any $\Psi_1 \perp \Psi_0$; (iii) $A_i^l$ has Schmidt rank $2^{\tilde O(1/\epsilon^3)} D^l$ across the cut $i|i+1$; (iv) $D\Delta \leq 1/2$ with $D = 2^{\operatorname{poly}\log(1/\epsilon)}$.*

We can get $D^2 \Delta \leq 1/2$ by modifying some unimportant constants in the construction of $A_i$.

**Definition 1.** *A state $\psi_l \in \mathcal{H}_{[1,i]}$ is a $(i, \delta, b)$-left state if (i) there exists a state $\psi_r \in \mathcal{H}_{[i+1,n]}$ such that $|\langle\Psi_0, \psi_l \otimes \psi_r\rangle| \geq \delta$; (ii) $\psi_l$ is an MPS of bond dimension $b$.*

**Definition 2.** *Let $\psi = \sum_{j\geq 1} \lambda_j l_j \otimes r_j$ be the Schmidt decomposition of a state across the cut $i|i+1$, where the Schmidt coefficients are in descending order: $\lambda_1 \geq \lambda_2 \geq \cdots > 0$. Define $\operatorname{trunc}_i^D \psi = \sum_{j=1}^D \lambda_j l_j \otimes r_j / \sqrt{\sum_{i=1}^D \lambda_i^2}$.*

The next lemma is an immediate corollary of the fact that the best rank-$D$ approximation to $\psi$ is $\operatorname{trunc}_i^D \psi$.

**Lemma 4** ([7]). *Suppose $\phi$ is a state of Schmidt rank $D$ across the cut $i|i+1$. Then, $|\langle \operatorname{trunc}_i^{D/\eta} \psi, \phi\rangle| \geq |\langle\psi,\phi\rangle| - \eta$ for any $\eta > 0$.*

The algorithm proceeds by iteratively constructing an $(i, \delta, b)$-left state for $i = 1, 2, \ldots, n$, where $b = n^{1+o(1)}/\delta$ with $\delta$ to be specified later. Each iteration has one random step that succeeds with probability $1 - O(n^{-2})$. Thus, the overall failure probability is $O(n^{-1})$.

Suppose we have an $(i-1, \delta, b)$-left state $\psi_l$. By definition 1, there exists a state $\psi_r = \sum_{j=1}^d \lambda_j |j\rangle_i \otimes r_j$ such that $|\langle\Psi_0, \psi_l \otimes \psi_r\rangle| \geq \delta$, where $\{|j\rangle_i\}_{j=1}^d$ is the computational basis of $\mathcal{H}_i$ and $r_j \in \mathcal{H}_{[i+1,n]}$. Lemma 3 implies that $A_i^l$ can be decomposed as $A_i^l = \sum_{J=1}^{2^{\tilde O(1/\epsilon^3)} D^l} \Lambda_J L_J \otimes R_J$, where $L_J$ ($R_J$) is a matrix product operator of bond dimension $2^{l \tilde O(1/\epsilon)}$ acting on $\mathcal{H}_{[1,i]}$ ($\mathcal{H}_{[i+1,n]}$). We have

$$\phi_1 \propto A_i^l \psi_l \otimes \psi_r = \sum_{J=1}^{2^{\tilde O(1/\epsilon^3)} D^l} \sum_{j=1}^d \Lambda_J \lambda_j L_J \psi_l |j\rangle_i \otimes R_J r_j,$$

$$|\langle\Psi_0, \phi_1\rangle|^2 \geq \delta^2/(\delta^2 + \Delta^l(1-\delta^2)) \geq 1 - \Delta^l/\delta^2. \qquad (1)$$

Let $\phi_2$ be a random state in $\text{span}\{L_J \psi_l \otimes |j\rangle_i\}$. With probability $1 - O(n^{-2})$, there exists $\psi'_r \in \text{span}\{R_J r_j\}$ such that $|\langle \phi_1, \phi_2 \otimes \psi'_r \rangle| \geq \Omega(n^{-2} D^{-l/2})$. Hence,

$$|\langle \Psi_0, \phi_2 \otimes \phi'_r \rangle| \geq \Omega(n^{-2} D^{-l/2}) - O(\Delta^{l/2}/\delta). \quad (2)$$

Choosing $l = O(\log n)$ and $\delta = n^{\text{poly} \log(1/\epsilon)}$ suitably, we have $\Delta^l = \Theta(\delta^4)$ and that the right-hand side of (2) is greater than $4\delta$.

We obtain $\phi_3$ by truncating each bond (in whatever order) of $\phi_2$ to $n^{1+o(1)}/\delta$. Lemma 1 implies an MPS $\Psi$ of bond dimension $n^{o(1)}$ such that $1 - |\langle \Psi, \Psi_0 \rangle| \ll \delta^2$. Hence $|\langle \Psi, \phi_2 \otimes \phi'_r \rangle| \geq 3\delta$. Lemma 4 implies $|\langle \Psi, \phi_3 \otimes \phi'_r \rangle| \geq 2\delta$. Finally, $|\langle \Psi_0, \phi_3 \otimes \phi'_r \rangle| \geq \delta$, and $\phi_3$ is an $(i, \delta, b)$-left state.

The final output of the algorithm is $\phi_1$ in the last iteration with the error estimate $|\langle \Psi_0, \phi_1 \rangle| \geq 1 - O(\delta^2)$. It is an MPS of bond dimension $b 2^{l \tilde{O}(1/\epsilon)} = n^{\tilde{O}(1/\epsilon)}$. It is easy to see that the running time of the algorithm is $n^{\tilde{O}(1/\epsilon)}$.

We acknowledge funding provided by the Institute for Quantum Information and Matter, an NSF Physics Frontiers Center (NSF Grant PHY-1125565) with support of the Gordon and Betty Moore Foundation (GBMF-12500028). In May 2015, Zeph Landau announced his joint work with Itai Arad, Umesh Vazirani, and Thomas Vidick on related problems in a talk [6] at the Simons Institute for the Theory of Computing. From the talk, the connection between their approach and the one in the present paper was not clear to me. Thanks to the correspondence with ZL in mid-October 2015, I realized that my approach has some overlap with theirs. Although the present work is independent, IA, ZL, UV, & TV deserve the credit of the overlapping part because they had the ideas since May 2015. They will post their results soon.

---

* ychuang@caltech.edu

[1] I. Arad, A. Kitaev, Z. Landau, and U. Vazirani. An area law and sub-exponential algorithm for 1D systems. arXiv:1301.1162v1.

[2] I. Arad, Z. Landau, and U. Vazirani. Improved one-dimensional area law for frustration-free systems. *Physical Review B*, 85(19):195145, 2012.

[3] C. T. Chubb and S. T. Flammia. Computing the degenerate ground space of gapped spin chains in polynomial time. arXiv:1502.06967.

[4] S. Gharibian, Y. Huang, Z. Landau, and S. W. Shin. Quantum Hamiltonian complexity. *Foundations and Trends in Theoretical Computer Science*, 10(3):159–282, 2015.

[5] Y. Huang. A polynomial-time algorithm for the ground state of one-dimensional gapped Hamiltonians. arXiv:1406.6355.

[6] Z. Landau. Local algorithms for 1D quantum systems. https://simons.berkeley.edu/talks/zeph-landau-2015-05-05. The slides were posted in mid-October 2015.

[7] Z. Landau, U. Vazirani, and T. Vidick. A polynomial time algorithm for the ground state of one-dimensional gapped local Hamiltonians. *Nature Physics*, 11(7):566–569, 2015.

[8] T. J. Osborne. Hamiltonian complexity. *Reports on Progress in Physics*, 75(2):022001, 2012.